\documentclass[aip,apl,reprint]{revtex4-1}

\usepackage{graphicx}

\begin{document}

\title{Spin Caloritronics in graphene with Mn} 

\author{Alberto Torres}
 \email{atriera@if.usp.br}
\author{Matheus P. Lima}
 \email{mplima@if.usp.br}
\author{A. Fazzio}
 \email{fazzio@if.usp.br}
\affiliation{Instituto de F\'isica de Universidade de S\~ao Paulo, CP 66318, 05315-970, S\~ao Paulo, SP, Brazil.}
\author{Ant\^onio J. R. da Silva}
 \email{ajrsilva@if.usp.br}
\affiliation{Instituto de F\'isica de Universidade de S\~ao Paulo, CP 66318, 05315-970, S\~ao Paulo, SP, Brazil.}
\affiliation{Laborat\'orio Nacional de Luz S\'incrotron, CP 6192, 13083-970, Campinas, SP, Brazil.}

\date{\today}

\begin{abstract}
We show that graphene with Mn adatoms trapped at single vacancies feature spin-dependent Seebeck effect, 
thus enabling the use of this material for spin caloritronics. 
A gate potential can be used to tune its thermoelectric properties in a way it presents 
either a total spin polarized current, flowing in one given direction, or currents for both spins flowing in 
opposite directions without net charge transport. Moreover, we show that the
thermal magnetoresistance can be tuned  between -100\% and +100\% by varying a
gate potential. 
\end{abstract}



\maketitle

The field of spin caloritronics deals with the interaction of spin and heat currents, that is, the coupling between 
spintronics and thermoelectrics\cite{BauerReview}.
Graphene\cite{Novoselov04} is a potential candidate material for spintronic devices
\cite{Graph-Spin-Filter,Graph-Spintronics-Nat} due to its long mean free path and weak spin-orbit coupling. 
Furthermore, it has been shown that graphene has potential for thermoelectric devices, both theoretically
\cite{Giant-S-Dragoman,zhou:073712} and experimentally\cite{TEP-Graphene-PRL,PhotoTEP-Perspective,Heer-TEP-graph}. 
Therefore, at least in principle, graphene is a good candidate for spin caloritronics.
Devices made of graphene nanoribbons with zigzag edges have already been proposed to this end
\cite{GNR-SpinCal,GNR-SpinCal2}.
Bi-dimensional graphene itself, although a good conductor for spin polarized currents, is spin degenerate, so it cannot 
be used as a \emph{source} for spin polarized currents. 
One way to lift its spin degeneracy that has been investigated recently is to dope graphene with metal adatoms
\cite{Krash-Metals-Graph,Matheus,Metals-Graph1,Metals-Graph2,Metals-Graph3,Metals-Graph4}.
Although their tendency is to diffuse and to form clusters, they can be trapped in defects like single vacancies (SV), 
where they are highly stable\cite{Matheus,Trap-Metal-Graph-JCP}.
SV defects in graphene can be created by ion\cite{ion-Beam-control} or electron\cite{Krash-Metals-Graph} irradiation, 
and, in the latter case, the use of focused beams allows sub-nanometer spatial control\cite{e-Beam-1Ang}.

\begin{figure}
 \centering
 \includegraphics[width=\linewidth]{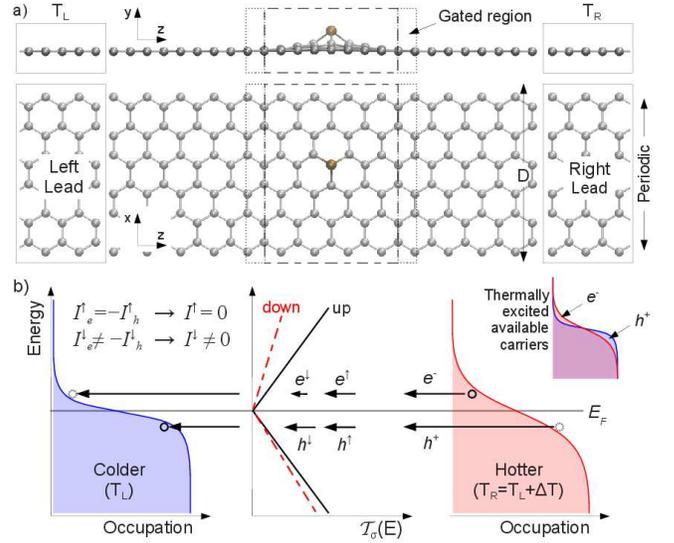}
 \caption{(Color online) (a) Geometry of the system studied: a graphene sheet (periodic in the $x$ direction)
 with a Mn atom trapped at a single vacancy, coupled to pristine graphene contacts.
 The central region is under the effect of gate potential.
 (b) Schematic illustration of the spin-dependent Seebeck effect mechanism.
 In the case shown, $\mathcal{T}$ is symmetric only for the up spin, leading
 to a net current only for the down spin channel.
 \label{fig: Geom}}
\end{figure}

It has already been shown by some of us that spin polarized currents appear in graphene doped with transition 
metals if spin-split localized levels that strongly hybridize with the $\pi$-bands of graphene are present 
close to the Fermi level ($E_f$)\cite{Matheus}. 
One particular feature that distinguishes the $Mn$ atom trapped in
SV (Mn@SV) from other configurations is its particular band structure, 
where there are occupied states with majority spin, and unoccupied states with minority
spin, both almost symmetrically positioned with respect to
$E_f$\cite{Matheus,Metals-Graph4,Metals-Graph1}. 
These localized states generate valleys with
majority (minority) spin below (above) $E_f$ in the transmission probability. Moreover, this transmission probability (and
thus the current) can be tuned by a gate potential\cite{Matheus}.
 
In this work, we calculate the thermoelectric properties and the spin-polarized currents of Mn@SV aiming to 
investigate its suitability for spin caloritronics. 
We show that i) the ferromagnetic (F) alignment is energetically favorable for
Mn-Mn distances greater than
$23$ \AA{}, whereas the anti-ferromagnetic (AF) alignment occurs for shorter distances; 
ii) it is possible to turn the current for a given spin channel arbitrarily small, 
or to have the up- and down-spin currents canceling each other by varying the
gate voltages ($V_g$);
iii)  the thermal magnetoresistance can be tuned to any desirable (from $-100\%$ to
$+100\%$) value by changing
$V_g$. 

Our calculations have been performed with non-equilibrium Green's functions coupled to density functional theory (NEGF+DFT). 
First, the geometries, comprised of a Mn@SV in a graphene supercell ($D\times20.0\times32.5\text{ \AA{}}^{3}$), were 
fully relaxed with a force criterion of 0.02 eV/\AA{} using the \textsc{siesta} code\cite{siesta}.
Here, $D$ is the distance between a $Mn$ atom and its lateral periodic image, as shown in Fig. \ref{fig: Geom}a. 
Then, we calculated the transmittances using the \textsc{transampa2} code\cite{transampa1,*transampa2,*GPU}. 
The electrodes were considered to be semi-infinite pristine graphene sheets.
In all calculations we used the Perdew-Burke-Ernzerhof generalized gradient approximation\cite{PBE} for the 
exchange-correlation functional, norm-conserving pseudopotentials\cite{Troullier} and a double-$\zeta$ polarized basis. 
We used an energy mesh cutoff of 300 Ry and a $7\times1\times4$ $k$-point sampling, in the Monkhorst-Pack
\cite{Monkhorst-Pack} scheme, to integrate the Brillouin zone. 
For the electronic transport calculations we employed 200 $k_{\bot}$-points (3000 for the PDOS). 
The gate potential was simulated by adding a smooth electrostatic potential to the 
Hamiltonian in a finite region (a $xy$ slab) containing the Mn@SV in the self-consistent cycle, thus allowing screening 
effects by charge rearrangement\cite{Matheus,Gate-Pot}. 
A vacuum layer of 20 \AA{} was used to avoid spurious interactions between periodical images in the direction 
perpendicular to the graphene plane ($y$ in Fig. \ref{fig: Geom}a).

\begin{figure}
 \centering
 \includegraphics[width=\linewidth]{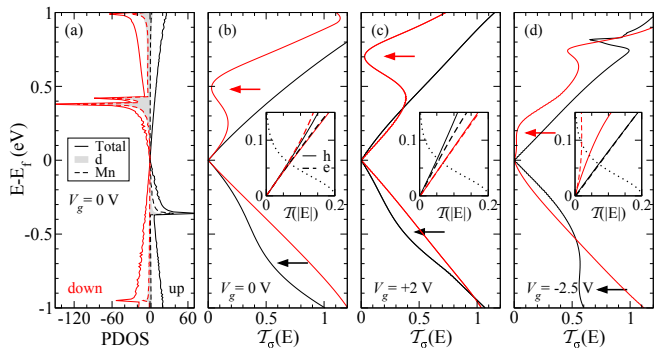}
 \caption{(Color online) (a) Projected density of states (PDOS) from Green's functions.
 (b-d) Spin resolved transmittance, $\mathcal{T}_{\sigma}(E)$, for gate voltages $V_g=0$, $+2$ and $-2.5$ V, 
 respectively.
 The insets show the asymmetry in $\mathcal{T}_{\sigma}(E)$ for electrons ($E>E_f$) and holes ($E<E_f$) near $E_f$.
 The dotted lines are $-\partial f/\partial E$ at 300 K (in arb. units). In all cases, $D=13$ \AA{}.
 \label{fig: TransDOS}}
\end{figure}

In the Landauer-B\"uttiker formalism, the current is given by\cite{Landauer-Buttiker,*Buttiker}
\begin{equation}
 I_{\sigma} = \frac{e}{h} \int \mathcal{T}_{\sigma}(E;V_{g}) \left( f_{L}-f_{R} \right) \, dE,
 \label{eq: I}
\end{equation}
where $\sigma=\uparrow,\downarrow$ is the spin, $e$ is the electron charge, $h$ is the Planck constant, 
$\mathcal{T}_{\sigma}(E;V_{g})$ is the spin resolved transmittance function, which depends on the gate voltage $V_{g}$;
$f_{L(R)}\equiv f(E,\mu,T_{L(R)})$ is the Fermi-Dirac distribution function, $\mu$ is the chemical potential of the 
electrodes and $T_{L(R)}$ is the temperature of the left(right) lead.

When the contacts are at different temperatures, the resultant unbalance in the density of thermally excited charge carriers,
given by $f_L-f_R$, allows electrons ($e^-$) and holes ($h^+$) to be available
to flow from the hot to the cold electrode, as shown in Fig. \ref{fig: Geom}b.
However, in order to the $e^-$ and $h^+$ currents not to cancel each other and a net current to be established, 
$\mathcal{T}_{\sigma}(E;V_{g})$ \emph{must} be asymmetric around $E_f$, that is, the transmittance for  $e^-$ 
($\mathcal{T}_{\sigma}^e$) and $h^+$ ($\mathcal{T}_{\sigma}^h$) needs to be different, 
otherwise $I_{\sigma}^e+I_{\sigma}^h=0$.

The Seebeck coefficient, also named thermoelectric power, is a measure of the voltage induced by a temperature 
difference and is defined as $S=-\Delta V/\Delta T|_{I=0}$. 
In the limiting case of $\Delta V \rightarrow 0$ and $\Delta T \rightarrow 0$ (the linear regime) an expression for $S$ 
can be derived from an expansion of Eq. (\ref{eq: I}), given by\cite{Imry-Seebeck}
\begin{equation}
 S_{\sigma} = -\frac{1}{e\overline{T}}
               \frac{\int \mathcal{T}_{\sigma}(E;V_{g}) \left(-\frac{\partial f}{\partial E}\right) (E-\mu) \, dE}
                    {\int \mathcal{T}_{\sigma}(E;V_{g}) \left(-\frac{\partial f}{\partial E}\right) \, dE},
 \label{eq: S}
\end{equation}
where $\overline{T}$ is the average temperature between the contacts.
Notice that the numerator of Eq. (\ref{eq: S}) (and therefore $S_{\sigma}$) is a measure of the local asymmetry of 
$\mathcal{T}_{\sigma}(E;V_{g})$ around $\mu$ in an energy range given by  $\partial f/\partial E$ (whose width, in 
turn, is determined by  $k_B\overline{T}$). 
We also see from Eq. (\ref{eq: S}) that if $\mathcal{T}_{\sigma}^e(E)$ is greater (smaller) than 
$\mathcal{T}_{\sigma}^h(E)$ (near $E_f$) the resultant $S_{\sigma}$ will be negative (positive).

In Fig. \ref{fig: TransDOS}a we show the projected density of states (PDOS) for the whole system, for the Mn atoms, 
and for the $d$ orbitals ($V_g=0$ and $D=13$ \AA{}).
As can be seen, the Mn@SV shows occupied (empty) localized $d$ levels nearly symmetrically located at approximately 
$-(+)0.4$ eV from $E_f$ for the $\uparrow$ ($\downarrow$) spins.
These levels give origin to valleys in $\mathcal{T}_{\sigma}(E)$, indicated by arrows in Fig. \ref{fig: TransDOS}b, 
which also affect the slope of the transmittance near $E_f$.
Positive $V_g$ (Fig. \ref{fig: TransDOS}c) raises the $d$ levels, moving the $\uparrow$($\downarrow$) spin valleys 
towards (away from) $E_f$.
On the other hand, negative $V_g$ (Fig. \ref{fig: TransDOS}d) lowers them, moving the $\uparrow$($\downarrow$) spin 
valleys away from (towards) $E_f$.
In the insets of Figs. \ref{fig: TransDOS}b-d, the asymmetry in $\mathcal{T}_{\sigma}(E)$ between $e^-$ ($E>E_f$) and 
$h^+$ ($E<E_f$) close to $E_f$ is shown in more detail.
For $V_g=0$ both spin channels are symmetric for $|E|\lesssim0.05$ eV. However, for $|E|\gtrsim0.05$ eV, 
$\mathcal{T}_{\uparrow}^e>\mathcal{T}_{\uparrow}^h$ (resulting in $S_{\uparrow}<0$) and 
$\mathcal{T}_{\downarrow}^e<\mathcal{T}_{\downarrow}^h$ (resulting in $S_{\downarrow}>0$). 
It is also interesting to note that, in this case, 
$\mathcal{T}_{\uparrow(\downarrow)}^{e(h)} \approx \mathcal{T}_{\downarrow(\uparrow)}^{h(e)}$, which means that 
(approximately) there is not a net charge transport because $I^{e}$ cancels
with $I^{h}$, but there \emph{is} spin transport 
because this canceling out is not from charge carriers of the same spin.
For $V_g=+2$ V, the $\downarrow$ spin valley is shifted away from $E_f$, rendering $\mathcal{T}_{\downarrow}^{e(h)}$ 
symmetric, and then, $S_{\downarrow}=0$. The $\uparrow$ spin valley is closer to
$E_f$, diminishing $\mathcal{T}_{\uparrow}$ for $E<E_f$, which makes
$\mathcal{T}_{\uparrow}^e>\mathcal{T}_{\uparrow}^h$, and thus $S_{\uparrow}<0$ (this case is similar to 
the one depicted in Fig. \ref{fig: Geom}b).
An opposite behavior happens for $V_g=-2.5$ V: the $\uparrow$ spin valley moves
away from $E_f$, rendering $\mathcal{T}_{\uparrow}$ symmetric between $e^-$ and 
$h^+$ (giving $S_{\uparrow}=0$); whereas the $\downarrow$ 
spin valley moves closer to $E_f$, yielding $\mathcal{T}_{\downarrow}$ highly asymmetric, 
lowering $\mathcal{T}_{\downarrow}^e$ in comparison to 
$\mathcal{T}_{\downarrow}^h$ (resulting in $S_{\downarrow}>0$).

\begin{figure}
 \centering
 \includegraphics[width=\linewidth]{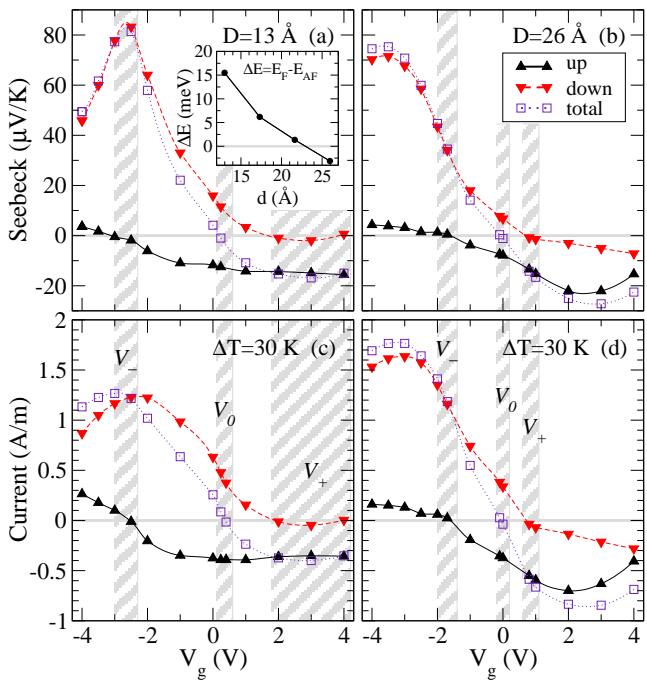}
 \caption{(Color online) Seebeck coefficient versus gate potential, $V_{g}$, for (a) $D=13$ \AA{} and (b) $D=26$ \AA{} at
 room temperature ($T_{L}=T_{R}=300$ K).
 Current versus $V_{g}$ for (c) $D=13$ \AA{} and (d) $D=26$ \AA{} ($T_{L}=300$ K and $T_{R}=330$ K).
 The inset in (a) shows the total energy difference between the F and the AF configurations, $\Delta E$, as a function of
 the Mn-Mn distance, $d$.
 The ranges $V_-$, $V_0$ and $V_+$ are discussed in the text.
 The lines are just guides to the eyes.
 \label{fig: S-IxV}}
\end{figure}

The total energy difference between the ferromagnetic and the anti-ferromagnetic alignment as a function of the Mn-Mn 
distance $d$ is shown in the inset of Fig. \ref{fig: S-IxV}a. 
To perform this simulation we laterally duplicate the geometry presented in Fig. \ref{fig: Geom}a, 
obtaining a supercell with two Mn@SV and $D=2d$. 
For $d \lesssim 23.0$ \AA{} the AF configuration is energetically favorable, indicating that a magnetic field is 
necessary to obtain the F configuration. 
On the other hand, for $d \gtrsim 23.0$ \AA{} the F alignment is the most favorable one. 
Note that only the F alignment has the spin unbalancing required to generate spin-polarized currents. 
Thus, all results presented in this work, except the thermally induced magnetoresistance, consider this magnetic alignment.

In Fig. \ref{fig: S-IxV} we show how the Seebeck coefficient and the thermally induced current vary with $V_g$ for $D=13$ 
and $26$ \AA{}.
For the larger $D$, the dispersion (and the broadening) of the localized $d$ levels get smaller, resulting in narrower valleys
in $\mathcal{T}_{\sigma}(E)$\cite{Matheus}. 
However, the qualitative behavior is very similar for both cases. 
As discussed before, $V_g$ shifts the valleys in $\mathcal{T}_{\sigma}$, which
allows one to tune the asymmetry in 
$\mathcal{T}_{\sigma}^{e(h)}$ and thus, to tune $S_\sigma$ and $I_\sigma$. 
From Fig. \ref{fig: S-IxV} it can be easily seen that there are three ranges of
$V_g$ (that we name $V_-$, $V_0$ and $V_+$), where  
the system can be tuned into three distinctive behaviors: 
(i) at $V_g \approx V_-$, $S_\uparrow \approx 0$ and $S_\downarrow>0$ ($\mathcal{T}_{\uparrow}^e \approx \mathcal{T}_{\uparrow}^h$ 
and $\mathcal{T}_{\downarrow}^e<\mathcal{T}_{\downarrow}^h$, see Fig. \ref{fig: TransDOS}d); 
$I_\uparrow \approx 0$ and there is only $I_\downarrow$ flowing; 
(ii) close to zero gate, for $V_g \approx V_0$, $S_\uparrow \approx -S_\downarrow$ 
($\mathcal{T}_{\uparrow}^{e(h)} \approx \mathcal{T}_{\downarrow}^{h(e)}$, see Fig. \ref{fig: TransDOS}b); in this case the system 
presents counter propagating spin currents without net charge transport; and, 
(iii) at $V_g \approx V_+$, $S_\uparrow<0$ and $S_\downarrow \approx 0$ ($\mathcal{T}_{\uparrow}^e>\mathcal{T}_{\uparrow}^h$ 
and $\mathcal{T}_{\downarrow}^e \approx \mathcal{T}_{\downarrow}^h$, see Fig. \ref{fig: TransDOS}c); 
$I_\downarrow \approx 0$ and there is only $I_\uparrow$ flowing.
For $D=13\,(26)$ \AA{}, these $V_g$ ranges are depicted in Fig. \ref{fig: S-IxV}.

\begin{figure}
 \centering
 \includegraphics[width=\linewidth]{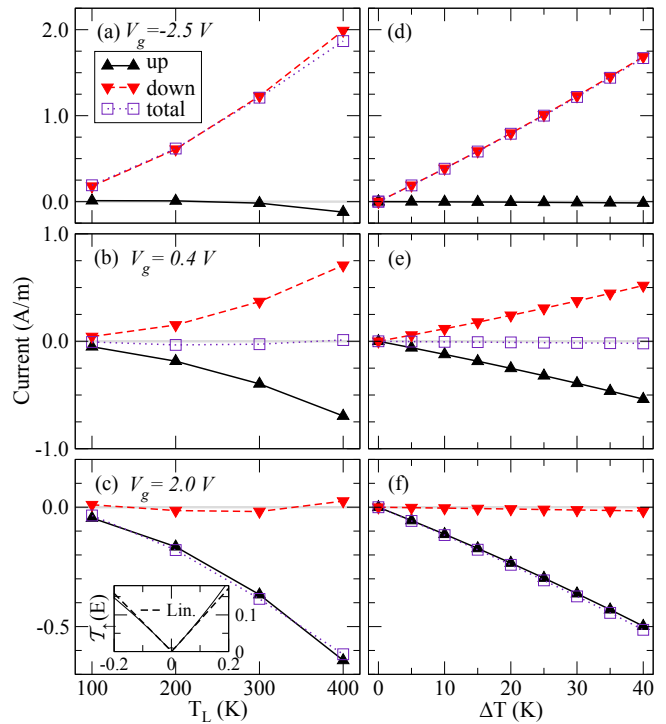}
 \caption{(Color online) Current versus (a-c) temperature, $T_L$, for constant $\Delta T=30$ K; and versus $\Delta T$ for
 $T_{L}=300$ K ($T_{R}=T_{L}+\Delta T$).
 Notice the different scales for the current. The lines are just guides to the eyes.
 \label{fig: IxT}}
\end{figure}

To investigate if the behaviors discussed above are robust under temperature changes, we calculated the 
current at different $T$ and $\Delta T$ for particular values of $V_g$ within $V_-$, $V_0$ and $V_+$ with $D$=13.0 \AA{}. 
In Figs. \ref{fig: IxT}a-c we show $I_\sigma$ as a function of $T_L$ for constant $\Delta T=30$ K, 
and in Figs. \ref{fig: IxT}d-f as a function of $\Delta T$ for fixed $T_L=300$ K. 
The temperature of the right contact was always varied as $T_R=T_L+\Delta T$.
When $T_L$ raises (constant $\Delta T$), $f_L-f_R$ broadens and some charge carriers are excited to higher energies.
This, combined with the deviation from linearity of $\mathcal{T}_{\sigma}(E;V_g)$ for $|E|$ high enough (see the inset
in Fig. \ref{fig: IxT}c) results in the non linear trend of $I_\sigma$ with $T_L$.
When $\Delta T$ varies (fixed $T_L$), the width of $f_L-f_R$ remains nearly constant but its amplitude grows linearly, 
giving $I_\sigma \propto \Delta T$.
Thus, the gate voltages necessary to achieve the situations (i),(ii) and (iii) are robust under changes in the temperature.

\begin{figure}
 \centering
 \includegraphics[width=\linewidth]{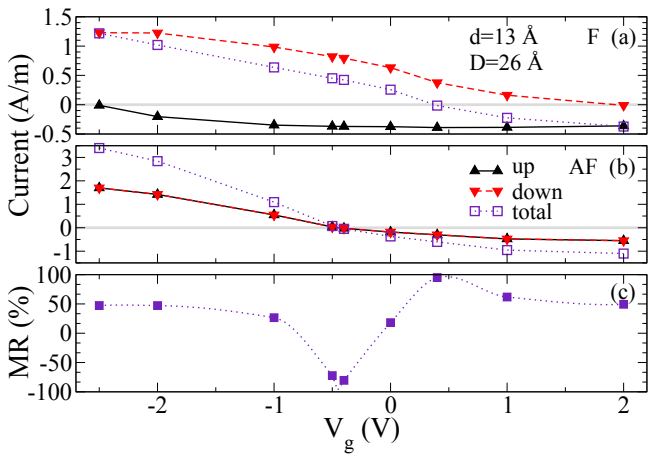}
 \caption{(Color online) Current of a system with two Mn@SV in the (a) ferromagnetic and (b) antiferromagnetic
 configurations as function of the gate voltage, $V_{g}$. (c) Thermal magnetoresistance.
 The lines are just guides to the eyes.
 \label{fig: F-AF-MRxV}}
\end{figure}

We also calculated the thermoelectric currents for the F and AF alignments, shown in
Fig. \ref{fig: F-AF-MRxV}a and Fig. \ref{fig: F-AF-MRxV}b, respectively. The F
alignment exhibits a pronounced spin-polarization in the current, whereas for
the AF configuration there is no spin-polarization in the current, as expected
due to the spin degenerated spectrum.

The thermally induced magneto-resistance (MR) is given by:
\begin{eqnarray}
MR [\%] =\frac{|I^F|^{-1}-|I^{AF}|^{-1}}{|I^F|^{-1}+|I^{AF}|^{-1}} \times 100.
\end{eqnarray}
This quantity depends only on the total currents
$I^{F(AF)}=I^{F(AF)}_\uparrow+I^{F(AF)}_\downarrow$ for the F (AF) alignments. 
Colossal MR are obtained when either $I^{F}$ or $I^{AF}$ are approximately zero. 
As seen in Fig\ \ref{fig: F-AF-MRxV}, in this system it is possible to obtain
both situations by varying $V_g$. For $V_g\approx 0.4$~V
$I^{F}_\uparrow=-I^{F}_\downarrow$,  leading to $I^F=0$, whereas 
for $V_g\approx -0.4$~V, $I^{AF}_\uparrow=I^{AF}_\downarrow=0$, leading to
$I^{AF}=0$. Thus, by varying $V_g$ within this region it is possible to control
the MR to any desirable value between -100\% and
+100\%, as shown in Fig. \ref{fig: F-AF-MRxV}c. 

Summarizing, the peculiar electronic structure of Mn@SV, with up and down states
positioned almost symmetrically with respect to the Fermi level allows a high
flexibility of the spin-dependent thermoelectric properties. The electron-hole asymmetry of
$\mathcal{T}_{\sigma}(E)$, and consequently the spin dependent Seebeck coefficient $S_\sigma$ can
be controlled by a gate voltage ($V_g$), leading to a suitable system for usage
in spin caloritronics.

\begin{acknowledgments}
We would like to thank A. R. Rocha for a critical reading of an earlier version of this manuscript. 
This work has received financial support from the brazilian agencies CNPq, INCT-Nanomateriais de Carbono and FAPESP. 
The calculations have been performed at CENAPAD-SP.
\end{acknowledgments}

\bibliography{References}

\begin{thebibliography}{31}
\bibitem [{GPU()}]{GPU}%
  \BibitemOpen
  \href@noop {} {}\bibinfo {note} {The code has been modified to allow the use
  of graphical processing units (GPUs)}\BibitemShut {NoStop}%
\bibitem [{Gat()}]{Gate-Pot}%
  \BibitemOpen
  \href@noop {} {}\bibinfo {note} {This potential consists of a plateau,
  $U(z)=eV_{g}$ for $z_{0}<z<z_{1}$, represented by a dashed line in Fig.
  \ref{fig: Geom}a, with smooth sinusoidal steps with width $\tau$, indicated
  by the dotted lines in Fig. \ref{fig: Geom}a. $U(z)=0$ elsewhere. In the
  calculations we defined the gated region trough $z_{0}=11.24$ \AA{},
  $z_{1}=21.23$ \AA{} and $\tau=1.25$ \AA{}.}\BibitemShut {Stop}%
\end{thebibliography}
\end{document}